# Hierarchical Dimensionless Learning (Hi-π): A physics-data hybrid-driven approach for discovering dimensionless parameter combinations

Mingkun Xia [a,b,c], Haitao Lin [a,b,c], Weiwei Zhang [a,b,c*]

a. School of Aeronautics, Northwestern Polytechnical University, Xi'an 710072，China

b. International Joint Institute of Intelligent Fluid Mechanics, Northwestern Polytechnical University, Xi'an 710072, China

c. National Key Laboratory of Aircraft Configuration Design, Xi'an 710072, China

* Corresponding author. E-mail address: aeroelastic@nwpu.edu.cn

**Abstract**

Dimensional analysis provides a universal framework for reducing physical complexity and reveal inherent laws. However, its application to high-dimensional systems still generates redundant dimensionless parameters, making it challenging to establish physically meaningful descriptions. Here, we introduce Hierarchical Dimensionless Learning (Hi-π), a physics-data hybrid-driven method that combines dimensional analysis and symbolic regression to automatically discover key dimensionless parameter combination(s). We applied this method to classic examples in various research fields of fluid mechanics. For the Rayleigh-Bénard convection, this method accurately extracted two intrinsic dimensionless parameters: the Rayleigh number and the Prandtl number, validating its unified representation advantage across multiscale data. For the viscous flows in a circular pipe, the method automatically discovers two optimal dimensionless parameters: the Reynolds number and relative roughness, achieving a balance between accuracy and complexity. For the compressibility correction in subsonic flow, the method effectively extracts the classic compressibility correction formulation, while demonstrating its capability to discover hierarchical structural expressions through optimal parameter transformations.

# 1. Introduction

With the rapid development of experimental measurement techniques (such as particle image velocimetry, wind tunnel experiments, etc.) and high-performance numerical simulations, fluid mechanics research has accumulated a large amount of data, but these data are highly dimensional, large-scale, and show significant data redundancy due to the strong constraints of the Navier-Stokes equation, which not only increases the complexity of model analysis, but also affects the experimental and computational costs, and prediction efficiency [1-3] .

Dimensional analysis is a physical dimension reduction method based on dimensional invariance (i.e., the form of physical laws does not depend on the choice of unit system)[4]. Almost all physical laws can be expressed by fewer dimensionless numbers and more concise dimensionless relationships. In the face of such high-dimensional redundant data, this method can simplify complex physical problems, reduce experimental measurement costs, improve the interpretability of physics, and has the ability to extrapolate across scales. It is a core tool for theoretical research. Especially in fluid mechanics, dimensional analysis has successfully achieved parameter reduction and flow feature extraction. Dimensionless numbers such as the Reynolds number and the Mach number provide a theoretical basis for revealing the essential laws of physical phenomena such as laminar-turbulent transition and compressible flow. At the same time, in the study of turbulence theory, the Kolmogorov -5/3 law derived by dimensional analysis has established a classic paradigm for understanding the multi-scale characteristics of turbulence.

However, in some high-dimensional systems, classical dimensional analysis and scaling law theory not only require a long time of derivation and repeated verification, but also rely heavily on expert experience, which makes this method often difficult to break through when facing complex systems. At the same time, the derived dimensionless quantities may still be large and unclear, and we have not been able to effectively extract fewer dimensionless parameters with significant characterization capabilities [5]. For example, for the Rayleigh-Bernard convection, dimensional analysis can obtain four dimensionless quantities, which is a large number, and their internal relationships (such as scaling laws) and relative sensitivity are not yet clear [6].

With the development of data science and machine learning, the limitations have been effectively solved [7-9]. Inspired by the idea of supervised dimensionality reduction (the framework is shown in **Figure 1**), we found that the limitations of classical dimensional analysis are mainly due to two key factors: first, dimensional analysis fails to effectively use existing data to capture the low-dimensional manifold structure in high-dimensional data; second, the dimensional analysis lacks mapping constraints on output parameters, resulting in the extracted dimensionless parameters may not be the optimal parameter combinations. Therefore, we need to combine dimensional analysis and supervised dimensionality reduction methods to establish a more accurate and universal parameter reduction method to find a small number of parameter combinations that have a key impact on the output, more effectively extract implicit physical laws from the data, and improve the generalization of physical modeling.

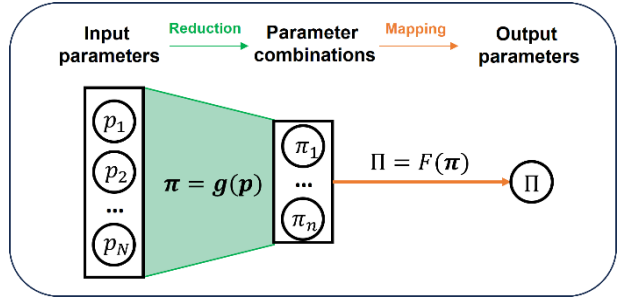

| Mapping / Reduction | Regression model | Neural network | Probability statistics |
|---|---|---|---|
| **Linear combination extraction** | Scaling LAWs [10] PyDimension [6] FIND[16] Active Subspaces [5,10-13] | Active Subspaces [14] NN_self-similarity [15] | IT-$\pi$[17] MIDL [18] |
| **Neural networks extraction** | —— | HiDeNN [19] BuckiNet [20] Conformal AE [21] | —— |
| **Symbolic regression extraction** | AST [22] SFL [23] | —— | —— |

**Figure 1 The idea of supervised dimensionality reduction (above).** Consider a system with $N$-dimensional input variables $\boldsymbol{p}$. The variable after dimensionality reduction is $\boldsymbol{\pi} = g(\boldsymbol{p})$, where the dimension of $\boldsymbol{\pi}$ is $n(1 \leq n < N)$, and $\boldsymbol{g}$ is the universal dimensionality reduction representation after the input parameters are de-redundant. At the same time, in order to meet the constraints of the target quantity, the variable $\boldsymbol{\pi}$ after dimensionality reduction satisfies $\Pi = F(\boldsymbol{\pi})$, where $F$ is the mapping relationship from the compression parameter $\boldsymbol{\pi}$ to the output parameter $\Pi$. **Classification table of data-driven dimensional analysis methods (below).** According to the idea of supervised dimensionality reduction, we split the existing methods into two levels for further classification. The first level is the dimensionality reduction from high-dimensional input parameters to low-dimensional parameter combinations (three categories), and the second level is the mapping of parameter combinations to output parameters (three categories). "——" means that we have not collected any work on this type of method.

In the research field of data-driven dimensional analysis, many methods have been proposed. We divide the supervised dimensionality reduction into two levels: reduction and mapping, as shown in **Figure 1**. The first-level reduction methods include linear combination extraction based on Buckingham Π theorem [5,6,10-18] (such as power-law form), nonlinear extraction based on neural networks (shallow neural network [19,20], autoencoder [21]), and nonlinear extraction based on symbolic regression [22,23]. The second-level mapping methods are mainly divided into two categories: one is model-based mapping, such as regression model class (linear regression [10], polynomial regression [6,23], Gaussian process regression [13], kernel function regression [22], etc.), neural network class [14,15,19-21], and the other is model-free mapping, such as probability statistics class (mutual information [17,18], etc.).

Next, we briefly summarize the above work from five directions and analyze its innovations and limitations.

**Invariance learning**: The first level of this direction converts the multiplication form of power exponents into a linear form for optimization extraction, and the second level uses different methods to characterize scaling law relationships and invariance. Mendez et al. proposed the Scaling LAWs, which combines dimensional homogeneity constraints with multivariate linear regression to discover physical power law relationships [10]. Xie et al. embedded the dimensional invariance principle into the first level of dimensionality reduction, and used a high-order polynomial model or a tree-based model in the second level to automatically identify the dominant dimensionless number of the system from experimental data using an alternating optimization strategy [6]. Watanabe et al. proposed a neural network-based method for discovering the self-similarity of physical systems from experimental or simulated data, and identified the power law exponent that characterizes the self-similar solution by optimizing network parameters [15]. Xiao et al. proposed FIND, which extended the first level of dimensionality reduction to the discovery of multiple dimensionless parameter combinations, which can solve more complex problems [16].

**Active subspace**: The first level of this direction is the same as invariance learning method, and the second level constructs a response surface to find the direction where the output parameters change significantly. Constantine et al. proposed to obtain active subspace through finite difference or response surface method, and then obtain unique or related dimensionless number combinations [5,10]. Jofre et al. applied this method to the heat transfer problem in irradiated particle turbulence and extracted important dimensionless combinations from the data [12]. Xu et al. applied artificial neural network to the construction of response surface and performed data-driven dimensional analysis to identify unique and related dimensionless quantities [14]. Zhang et al. divided the data through cluster analysis and extracted dimensionless combinations in different regions through active subspace method [13].

**Neural network**: The first level in this direction is represented by shallow networks or autoencoders, and the mapping relationship in the second level is generally represented by complex deep networks. Saha et al. proposed HiDeNN, which uses the BIC criterion to weigh the simplicity and accuracy of scaling networks, thereby discovering the most physically interpretable dimensionless parameters [19]. Bakarji et al. proposed BuckiNet, which converts the power multiplication form unique to dimensionless parameters into a shallow network, uses dimensionless constraints as the loss function of the shallow network, and directly extracts dimensionless combinations from dimensional quantities [20]. Evangelou et al. use autoencoders to compress and extract the main parameter combinations and redundant parameter combinations that affect output behavior [21].

**Symbolic regression**: The first level in this direction is represented by symbolic regression methods, and the mapping relationship in the second level is represented by simple polynomials or kernel function models. Luo et al. proposed the adaptive spatial transformation AST method, which uses symbolic regression based on genetic algorithms to extract the optimal parameter combination and perform extrapolation prediction [22]. Lin et al. proposed the scaling function learning SFL method based on the AST method. The loss function is the mean square error of the polynomial fitting, so as to extract generalizable parameter combinations [23].

**Probability statistics**: The first level of this direction is the same as invariance learning method. The mapping relationship of the second level is in the form of no explicit model. Mutual information is used to measure the influence relationship between parameter combinations and output parameters. Yuan et al. and Zhang et al. almost simultaneously proposed a dimensional analysis method based on mutual information, using mutual information (MI) as an objective criterion to identify key physical dimensionless quantities [17,18]. Yuan et al. automatically extracted the optimal dimensionless parameters, physical states and characteristic scales in a model-free manner, expanding its scope of application.

The above methods automatically identify the dominant parameter combinations from a large amount of experimental data or simulation results, providing a deep understanding of complex physical phenomena. However, these methods still have limitations to a certain extent. Methods such as invariance learning, active subspace, and probability statistics all assume that dimensionless parameter combinations exist in the form of power multiplication. This assumption limits their scope of application in complex scenarios. In the work of Luo et al. [22] and Lin et al. [23], symbolic regression methods were used to discover complex nonlinear dimensionless parameter combinations, breaking through the parameter representation in the form of power multiplication. Meanwhile, methods such as the dimensionless learning method proposed by Xie et al. [6] only extract a single dominant parameter combination, which will fail when extracting multiple parameter combinations. In order to solve this problem, we use the symbolic regression method to extract multiple parameter combinations through its multi-branch tree structure, and in most cases, the global optimal expression will be obtained. Neural network methods have powerful expressive power, but their training process is complex and hyperparameter tuning is difficult. Symbolic regression methods are significantly superior to black box modeling methods such as neural networks due to their simplicity, flexibility, and white box interpretable modeling.

Therefore, we propose Hi-π, a hierarchical dimensionless learning method, which aims to extract dimensionless parameter combinations with significant representation capabilities. This method is based on the idea of supervised dimensionality reduction and is divided into three layers: physical layer, data layer, and mapping layer. The first layer (physical layer) uses dimensional analysis to ensure physical constraints; in the second layer (data layer), we use symbolic regression method to reduce the dimension of variables, and the obtained parameter combination is not unique and does not need to be multiplied by power exponentials as the hypothesis of parameter combination; the third layer (mapping layer) uses multivariate polynomial regression to represent, ensuring that the model is smooth, global, and can characterize nonlinear relationships.

The rest of this paper is organized as follows. In the second part, we describe in detail the principle of the Hi-π method to extract dimensionless parameter combinations. In the third part, we will demonstrate the advantages of parameter reduction and knowledge discovery of Hi-π method through mathematical examples (Section 3.1), Rayleigh-Bernard convection example (Section 3.2), rough circular pipe flows example (Section 3.3), and compressibility correction example (Section 3.4). In the fourth section, we summarize the paper and discuss future research.

## 2. Hierarchical Dimensionless Learning (Hi-𝜋)

Based on dimensional analysis, we propose Hi-π, a hierarchical dimensionless learning method driven by physics (dimensional analysis) and data (symbolic regression and polynomial regression) to extract parameter combinations that have a key impact on the output, as shown in **Figure 2**. Next, we will present the details of the framework in three parts.

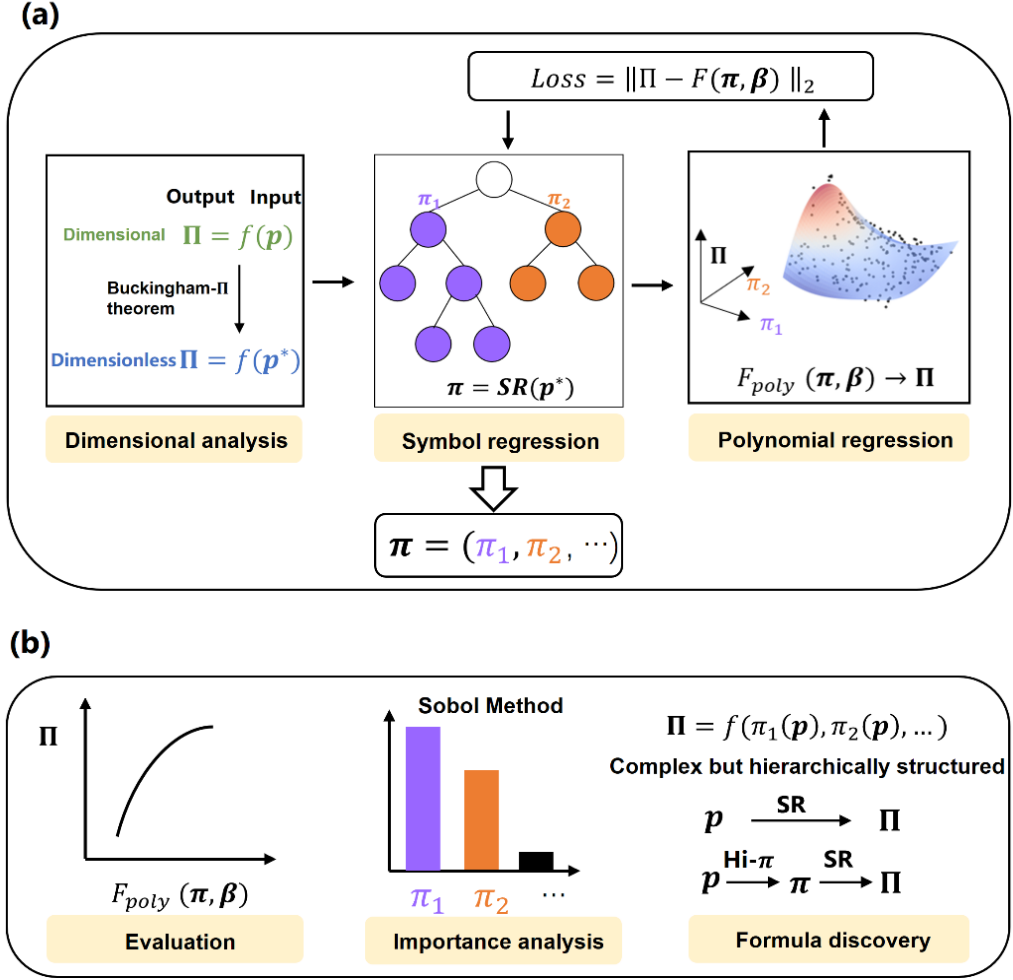

**Figure 2 Schematic diagram of hierarchical dimensionless learning (Hi-π)** (a) Method flow: three-stage hierarchical modeling strategy of dimensional analysis-symbolic regression-polynomial regression; (b) Evaluation application: sensitivity analysis and predictive evaluation of parameter combinations are performed. At the same time, parameter combinations can also be extracted and integrated into the mathematical expression discovery process, which can effectively improve the recognition accuracy.

### Step 1: Initialization

Problems in the field of physics such as fluid mechanics often involve many physical quantities, which requires us to accurately determine the problem variables and establish a clear input-output relationship in the first step. The output variable $\Pi$ is usually the core indicator of the research target, namely the system response quantity, which directly reflects the performance or behavior of the system; the input variable $p$ is the control quantity closely related to the system behavior, and these quantities jointly determine the change of the output variable.

Among them, the input parameters usually constitute a relatively high-dimensional parameter space, which increases the complexity of symbolic regression to extract parameter combinations. In order to simplify the problem and extract the key physical relationship, dimensional analysis becomes a powerful tool. We provide two ideas:
- **Dimensional embedding**: Through dimensional analysis, especially Buckingham Π theorem, we can combine multiple physical quantities into dimensionless parameters, and then use them as the input of symbolic regression, which can accelerate the extraction efficiency of parameter combinations. This method is suitable for when there are many input parameters. This paper mainly adopts this idea.
- **Dimensional constraint**: Introducing dimensional homogeneity constraints in the symbolic regression process can avoid generating expressions that do not conform to physical dimensions and improve the accuracy of parameter combinations. This method is suitable when there are relatively few input parameters. The PySR algorithm already provides this function [26].

**Step 2: Extraction**

Symbolic regression is a method that automatically discovers mathematical models from data and is widely used in scientific modeling and theoretical research [24,25]. In this step, we use the PySR symbolic regression method based on a multi-population evolutionary algorithm [26]. The algorithm in PySR uses a tree structure to cleverly represent complex mathematical expressions. At the same time, another advantage of this tree structure is that it can perform parallel search of expressions through a multi-branch tree structure to obtain multiple parameter combinations $\pi_i (i = 1,2,…,n)$. As shown in **Figure 2**, we divide the expression tree into two branches, which can simultaneously extract two parameter combinations $\pi_1$ and $\pi_2$. The number of parameter combinations can be tried and selected according to the specific problem.

In order to more effectively evaluate the parameter combinations obtained by symbolic regression, we need to construct a fitness function, namely the loss function. The loss function should be able to measure the predictive generalization ability of the parameter combination for the target quantity while ensuring the simplicity and universality of the expression. Multivariate polynomial fitting is a simple and universal method that is smooth, global, and can characterize nonlinear relationships. Therefore, it can be used to evaluate the prediction error between the parameter combination $\boldsymbol{\pi}$ and the output variable $\Pi$. The loss function can be expressed as:

$$Loss = \|\Pi - F(\boldsymbol{\pi}, \boldsymbol{\beta})\|_2 \qquad (1)$$

where $F(\boldsymbol{\pi}, \boldsymbol{\beta})$ is the predicted output value obtained by multivariate polynomial fitting, and is generally extracted using an order of 3-5. PySR continues to search for the optimal parameter combination through iterative operations such as inheritance, crossover, and mutation until the loss function converges. If there are simplifiable sub-parameters in the parameter combination extracted by symbolic regression, parameter dimension reduction can be further achieved through manual extraction.

**Step 3: Application**

The application process is divided into importance analysis, prediction evaluation and extension (integrating parameter combination extraction into the mathematical expression discovery process). The extension is described in Section 3.4.

In order to better evaluate the relative importance of multiple parameter combinations, we use the sensitivity analysis method of Sobol variance decomposition [27]. The core of the Sobol method is to decompose the total variance into the contribution of each input variable and its interaction term to the output variance. By decomposing the variance of the model output into different parts, the uncertainty contribution of each input variable or their combination to the output can be quantified. Its core formula is:

$$Var(\Pi) = \sum_{i=1}^{n} S_i + \sum_{1 \leq i < j \leq n} S_{ij} + \cdots \qquad (2)$$

$$S_i = \frac{Var[\mathbb{E}(\Pi|\pi_i)]}{Var(\Pi)} \qquad (3)$$

$$S_{ij} = \frac{Var[\mathbb{E}(\Pi|\pi_i, \pi_j)]}{Var(\Pi)} - S_i - S_j \qquad (4)$$

where $Var(\Pi)$ represents the total variance of the output variable $\Pi$, $S_i$ is the first-order sensitivity index, which represents the independent contribution of a single input variable $\pi_i$ to the total variance, $S_{ij}$ is the second-order sensitivity index, which represents the contribution of the interaction between input variables $\pi_i$ and $\pi_j$ to the total variance, and so on.

In addition to the above decomposition, the Sobol method also introduces the total effect sensitivity index $S_{Ti}$ to measure the total contribution of the input variable $\pi_i$ and all its interactions with other variables to the output variable $\Pi$. The calculation formula is

$$S_{Ti} = 1 - \frac{Var[\mathbb{E}(\Pi|\pi_{\sim i})]}{Var(\Pi)} \qquad (5)$$

where $\pi_{\sim i}$ represents all input variables except $\pi_i$.

## 3. Application

In this section, we use mathematical and physical examples to verify the effectiveness of hierarchical dimensionless learning (Hi-π) method and compare it with existing methods.

### 3.1 Mathematical Examples

To facilitate understanding while taking into account complexity, we design the following three representative mathematical examples to verify the algorithm, covering 1-3 parameter combinations, as shown in **Table 1**.

Table 1 Mathematical examples

| Example | Parameter combinations | Mathematical expressions and value ranges |
|---|---|---|
| 1 | $\sqrt{x_0} + x_1(x_2)^2$ | $f_1 = \ln(\sqrt{x_0} + x_1(x_2)^2)(\sqrt{x_0} + x_1(x_2)^2)^2$ <br> $(x_0, x_1, x_2) \in (0.5, 2.3)^3$ |
| 2 | $\sqrt{x_0} + x_1,\ x_0(x_2)^2$ | $f_2 = \ln(\sqrt{x_0} + x_1)(x_0(x_2)^2)^2$ <br> $(x_0, x_1, x_2) \in (0.5, 2.3)^3$ |
| 3 | $x_0 x_1 + x_3, \sqrt{x_2} + x_3, \dfrac{x_4^2}{x_5}$ | $f_3 = \exp(x_0 x_1 + x_3) + \ln(\sqrt{x_2} + x_3)\left(\dfrac{x_4^2}{x_5}\right)^2$ <br> $(x_0, x_1, x_2, x_3, x_4, x_5) \in (0.5, 1.5)^6$ |

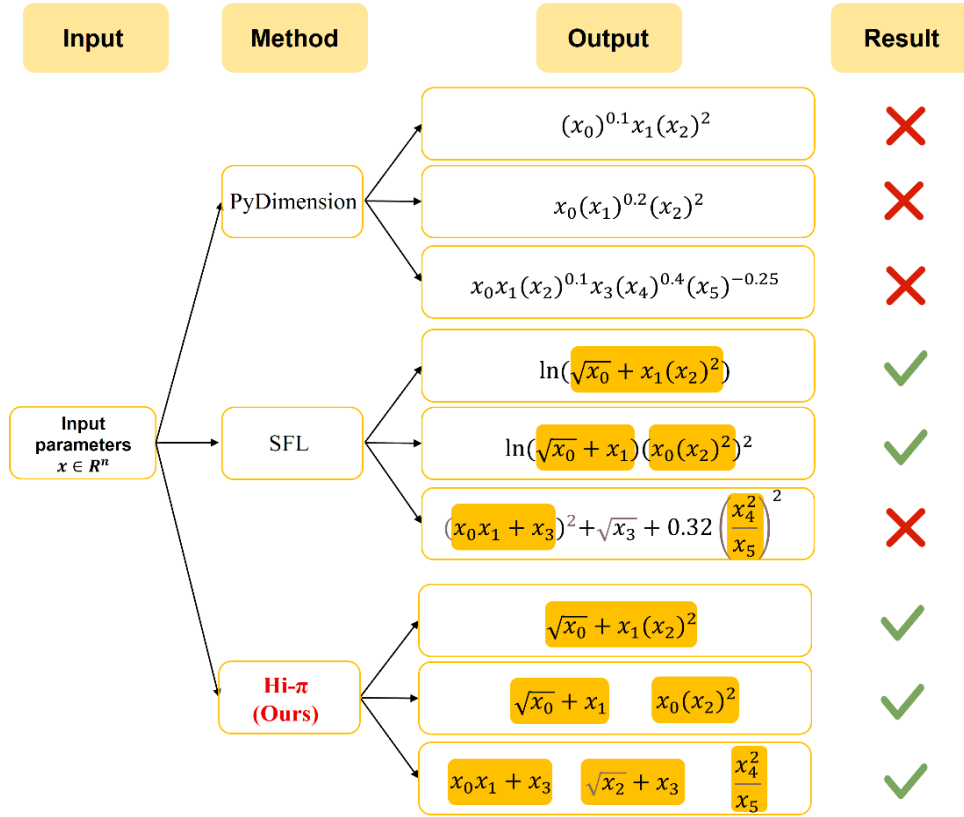

**Figure 3 Schematic diagram of the extraction parameter combination of three methods** (PyDimension method, SFL method and Hi-π method). The whole process from input to method to output, and finally the result evaluation, is demonstrated using three different mathematical examples (each method is example 1, 2, and 3 from top to bottom). The PyDimension method fails to extract, the SFL method can extract part of them, and the Hi-π method can extract all of them. For the two symbolic regression methods (SFL method and Hi-π method), we will then make a detailed comparison.

Since most of the parameter combinations in **Table 1** cannot be written in the form of power multiplication, most linear combination extraction methods based on Buckingham Π theorem may not be able to extract effectively, while symbolic regression methods are not restricted by the assumption and can obtain relatively more complex and flexible parameter combinations. In the example of multiple parameter combinations, the Hi-π method can directly output the expressions corresponding to multiple parameter combinations, while the SFL method can only obtain one expression. If these parameter combinations are embedded in the results obtained by the SFL method, then we believe that the SFL method can extract them, otherwise it cannot, as shown in **Figure 3**.

Next, we will compare the ability of two symbolic regression methods (i.e., Hi-π method and SFL method) to extract parameter combinations through three mathematical examples in **Table 1**, and prove the feasibility and noise robustness of the Hi-π method. In order to ensure the rigor of the comparison experiment between the two methods, we set the data initialization, hyperparameter selection, and noise level to be the same, as shown in **Table 2**. By selecting different random initial states and repeating them multiple times, the extraction accuracy of the Hi-π method and the SFL method is compared. After the comparative experiment, the final result is shown in **Figure 4**. **Figure 4** shows that in the presence of multiple parameter combinations, the Hi-π method can effectively extract, while the SFL method may completely fail. In addition, in a noisy data environment, the Hi-π method shows strong robustness, which further verifies its reliability in practical applications. In contrast, single parameter extraction methods such as the SFL method have obvious limitations.

**Table 2** Parameter settings

1. **Sample construction**: randomly sample 100 in the corresponding range
2. **Symbol library**: +, -, *, /, sqrt, square, inv, exp, log, sqrt, abs, cube, sin, cos
3. **Polynomial order**：5
4. **Noise level**: within 5%

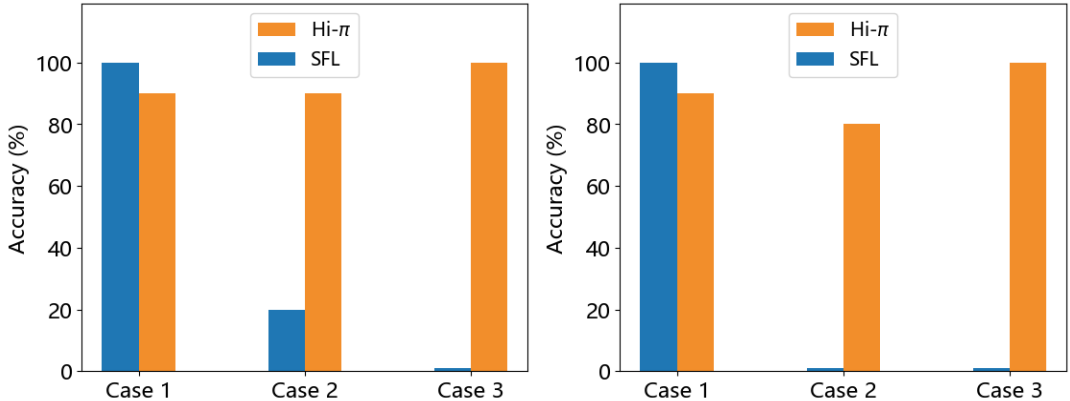

**Figure 4 The results of the comparative experiment**. The left image shows the experimental group without noise, and the right image shows the experimental group with noise.

In the comparative experiment, we found that the SFL method partially failed in Example 2 and completely failed in Example 3. We used the Sobol method to analyze the sensitivity of the corresponding parameter combinations, and found that there is a strong and weak sensitivity relationship between the parameter combinations, as shown in **Figure 5**. Under the parameter combination with strong sensitivity, the weak parameter combination will be approximately constant, so the SFL method can capture the strong parameter combination or is strong, and cannot extract the weak parameter combination, and will fall into the local optimal solution; the Hi-π method is not easily affected by this, can extract at the same time, and is more generalized when extrapolating.

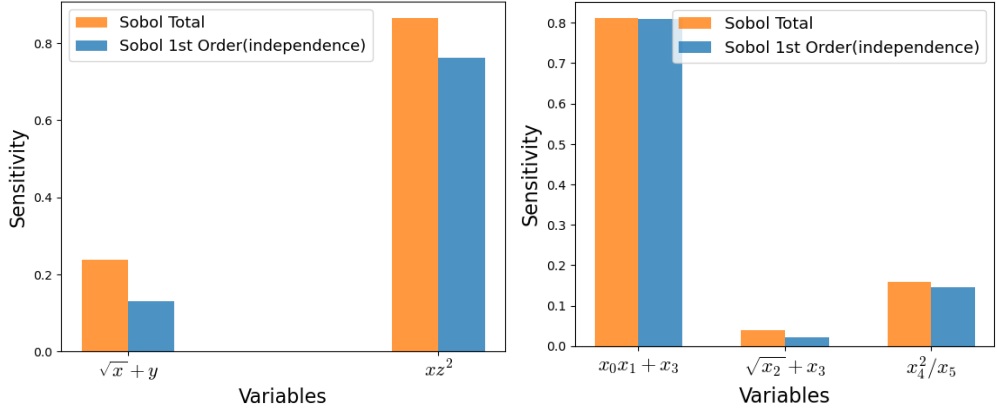

**Figure 5 Comparison of the sensitivity of parameter combinations.** The left figure shows the sensitivity analysis of parameter combinations in Example 2, and the right figure shows the sensitivity analysis of parameter combinations in Example 3. Sensitivity analysis proves the strong and weak relationship of parameter combinations.

### 3.2 Rayleigh-Bernard convection

Thermal convection is a turbulent motion phenomenon driven by density differences caused by temperature differences. It is widely present in nature, engineering, and life. A classical fluid mechanics model is abstracted from these complex phenomena, now known as Rayleigh-Bernard convection [28]. Rayleigh-Bernard convection studies the turbulence caused by buoyancy, viscosity, and gravity in the heated bottom and cooled top of a heated container. In this section, we will rediscover the main dimensionless parameter combination in this typical fluid system, namely the Rayleigh number Ra and the Prandtl number Pr.

The physical quantities included in this fluid mechanics model include: heat flow $q$, container height $h$, temperature difference $\Delta T$, gravitational acceleration $g$, and thermal conductivity $\lambda$, thermal expansion coefficient $\alpha$, kinematic viscosity $\nu$, and thermal diffusivity $\kappa$ of the fluid. For the Rayleigh-Bernard convection system, the core issue is how turbulent flow transports heat, that is, the law of change of the Nusselt number. The Nusselt number $Nu = \frac{qh}{\lambda \Delta T}$ measures the heat transfer efficiency of the system and can be used as the output parameter of the system. The input parameter $\boldsymbol{p}$ includes other physical quantities except $q$, namely $(h, \Delta T, \lambda, g, \alpha, \nu, \kappa)$. The following expression can accurately describe the relationship between all physical quantities:

$$Nu = \frac{qh}{\lambda \Delta T} = f(h, \Delta T, \lambda, g, \alpha, \nu, \kappa) = f(\boldsymbol{p}) \tag{6}$$

The above expression defines the causal relationship of the Rayleigh-Bernard convection system, in which there are seven input parameters. This is a relatively high-dimensional parameter space, which is not conducive to physical analysis and prediction. Through dimensional analysis, such as Buckingham Π theorem, we can combine multiple physical quantities into fewer dimensionless parameters to reduce the input variables of the expression. The dimensional matrix D of the Rayleigh-Bernard convection example can be expressed as

$$D = \begin{matrix} & h & \Delta T & \lambda & g & \alpha & \nu & \kappa \\ [L] & \begin{bmatrix} 1 & 0 & 1 & 1 & 0 & 2 & 2 \\ [T] & 0 & 0 & -3 & -2 & 0 & -1 & -1 \\ [M] & 0 & 0 & 1 & 0 & 0 & 0 & 0 \\ [\Theta] & 0 & 1 & -1 & 0 & -1 & 0 & 0 \end{bmatrix} \end{matrix} \tag{7}$$

where [L] [T] [M] [Θ] are the dimensions of length, time, mass, and temperature, respectively.

In order to satisfy the dimensional invariance constraint, the dimensional matrix D must satisfy

$$Dw_{bi} = 0 \; (i = 1,2,3) \tag{8}$$

In this example, there are infinite results that satisfy this condition, and we arbitrarily select three of them as basis vectors, namely

$$w_{b1} = [0,1,0,0,1,0,0]^T$$
$$w_{b2} = [0,1,0,0,1,1,-1]^T$$
$$w_{b3} = [3,0,0,1,0,-2,0]^T \tag{9}$$

So the causal expression (6) can be written as

$$Nu = f(s_1, s_2, s_3) = f(\alpha \Delta T, \frac{\nu \alpha \Delta T}{k}, \frac{gh^3}{\nu^2}) \tag{10}$$

The classic Buckingham Π theorem is a dimensionality reduction method based on physical laws and dimensional invariance. We can compress the input variables from seven to three, but three dimensionless parameters are still difficult for physical analysis. Since there are often low-dimensional structures inside high-dimensional data, we can use a data-driven method to further reduce the dimensionality of the 3 dimensionless parameters. The relationship after dimensionality reduction can be expressed as

$$Nu = f(\pi_1, \ldots, \pi_n), n \leq 3 \qquad (11)$$

where $\pi_n = g_n(s_1, s_2, s_3)$. Traditional dimensional analysis assumes that $g$ satisfies the form of power multiplication. Next, we use symbolic regression without formal constraints to extract it, and prove that this dimensionless parameter is indeed expressed in the form of a power law.

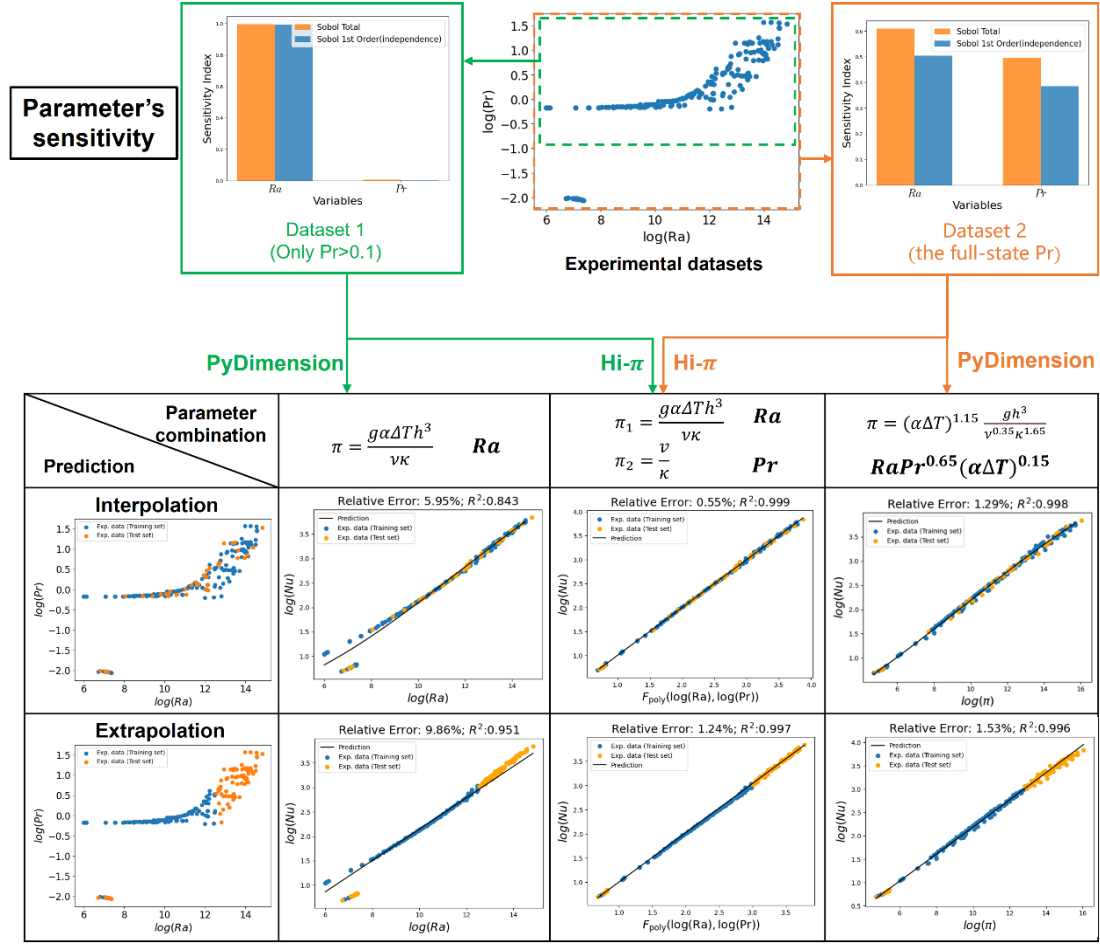

**Figure 6 Parameter combination obtained by two methods for Rayleigh-Bernard convection example, their prediction comparison, and parameter sensitivity**. The sensitivity of dimensionless parameters in different data ranges is different, which also leads to different results extracted by PyDimension method, while the Hi-π extraction results are completely consistent, which are the two intrinsic dimensionless parameters of Rayleigh number and Prandtl number. In the prediction comparison part, the Hi-π extraction results have the highest accuracy regardless of interpolation or extrapolation, which further proves the ability of Hi-π to extract intrinsic dimensionless parameters.

We collected Rayleigh-Bernard convection data from three different papers [29-31]. At the same time, in order to solve the problem of data consistency, we retained the data of turbulent state for analysis and deleted the incompatible transition state data. Then, according to the size of Prandtl number, we extracted the large Prandtl number subset (Dataset 1, Pr>0.1) from the original data set, and retained the complete original dataset (Dataset 2, containing the full range of Pr data). Then we used the Hi-π method to carry out further data-driven dimensionality reduction, and compared the results with the existing method PyDimension [9]. Both methods employed a fifth-order polynomial as the mapping layer. In the prediction comparison, the interpolation part selected 80% of the data as the training set and 20% of the data as the test set; the extrapolation part selected the range of lg(Ra)≤12.5 and Pr≥0.0095 as the training set. There were only three data of small Prandtl number in the training set, and the test set contained data of large Rayleigh number and small Prandtl number, which were more difficult to measure. The results are shown in **Figure 6**.

When the Prandtl number Pr>0.1(Dataset 1), the PyDimension method only obtains a single dominant dimensionless parameter, namely the Rayleigh number Ra. The Hi-π method extracts both the Rayleigh number Ra and the Prandtl number Pr. This result is more consistent with the theory. Ra measures the strength of buoyancy drive and viscous resistance, and determines whether convection occurs and its strength; Pr describes the relative speed of viscous diffusion and thermal diffusion, and affects the coupling of flow and temperature fields. Under the data of the full-state Prandtl number (Dataset 2), the parameter combination extracted by the Hi-π method is still the Rayleigh number Ra and the Prandtl number Pr, but the PyDimension result is different. Although the accuracy is also high, the robustness is not high. At the same time, in the prediction comparison link, we found that the parameter combination extracted by the Hi-π method has the highest accuracy in both interpolation and extrapolation, which further shows that the Hi-π method can capture the intrinsic parameter combination with significant characterization ability in the physical system. In **Figure 6**, we use the Sobol method to compare the sensitivity of the Rayleigh number and the Prandtl number. When the Prandtl number Pr>0.1, the dominant characteristic of the Rayleigh number is very obvious, so the PyDimension method can only find the Rayleigh number. However, under the full-state Prandtl number, the sensitivity of the Rayleigh number and the Prandtl number is equivalent, so the extraction results are inconsistent and affected by the parameter sensitivity, which is consistent with the Lindborg's conclusion [32]. The hierarchical dimensionless learning method we proposed is not affected by the data range and the sensitivity of the parameter combination, and can find the low-dimensional parameters inherent in the system itself.

In the symbolic regression method for extracting parameter combinations, the number of dimensionless parameter combinations is usually regarded as a known condition. However, for general problems, this number is often difficult to determine in advance. To this end, we can adopt an iterative strategy: first, assume that there is only one dominant dimensionless parameter in the system, and extract parameters based on the hierarchical symbolic regression method; then verify the rationality of this assumption by evaluating the model prediction error. If the single parameter assumption cannot meet the accuracy requirements, the number of branches in the tree structure in the PySR method is gradually increased to systematically increase the dimension of the parameter space. Through this gradual parameter expansion and verification process, the intrinsic dimension of the low-dimensional manifold of the complex physical system can be finally determined.

## 3.3 Viscous flows in a circular pipe

As the core part of industrial transportation systems, pipe flow has a direct impact on energy consumption, transportation efficiency, and operating costs. This makes the study of rough pipe flow of significant engineering application value. Therefore, the problem of rough pipe flow has always been an important topic in fluid mechanics' research. In a classic study, Nikuradse established an experimental database for the relationship between friction factor, Reynolds number, and relative roughness for the first time through the uniform particle coating technology on the pipe wall [33]. His research results not only provided an empirical formula for engineering resistance calculation, but also became a standard model for the application of dimensional analysis theory [34,35].

The physical quantities involved in the entire flow system include: pressure loss $\Delta p$, average velocity $V$, kinematic viscosity $v$, pipe diameter $D$ and pipe wall roughness $\varepsilon$. The pressure loss reflects the energy loss caused by the fluid overcoming friction and turbulence in the pipe, and is the dependent variable of the entire system. The Darcy friction factor $f_D$ can be obtained by non-dimensionalizing it, that is,

$$f_D = \frac{\Delta p \, D}{\frac{1}{2}\rho V^2 L} \tag{12}$$

Darcy friction factor is more commonly used to describe the overall resistance characteristics of the pipeline system and is the output parameter of the entire system. The input parameter p includes other physical quantities except the pressure loss $\Delta p$, namely $(V, v, D, \varepsilon)$. The following expression can accurately describe the relationship between all physical quantities:

$$f_D = f(V, v, D, \varepsilon) = f(\boldsymbol{p}) \tag{13}$$

where there are four input parameters. The dimensionality of the input variables is reduced by dimensional analysis. The dimensional matrix D of the rough pipe flow system can be expressed as

$$D = \begin{matrix} & V & v & D & \varepsilon \\ [L] & 1 & 2 & 1 & 1 \\ [T] & -1 & -1 & 0 & 0 \end{matrix} \tag{14}$$

where dimensions include length [L] and time [T].

In order to satisfy the dimensional invariance constraint, the dimensional matrix D must satisfy

$$Dw_{bi} = 0 \ (i = 1,2) \tag{15}$$

Buckingham Π theorem analysis reduces the input variables from four to two. In this example, there are infinitely many dimensions that satisfy the dimensionless condition $Dw_b = 0$, and the results we want are the Reynolds number Re and the relative roughness, which are already recognized dimensionless numbers. So how to select the best and most physically interpretable dimensionless number combination, that is, to rediscover the Reynolds number and relative roughness?

According to the research conclusions of classical dimensional analysis theory and engineering empirical models, when the Reynolds number $Re \leq 3000$, the friction factor $f_D$ follows the Poiseuille formula under laminar flow:

$$f_D = \frac{64}{Re} \tag{16}$$

When $Re > 3000$, the friction factor $f_D$ is described by the Colebrook implicit equation:

$$\frac{1}{\sqrt{f_D}} = -2log_{10}\left(\frac{1}{3.7}\frac{\varepsilon}{D} + \frac{2.51}{Re\sqrt{f_D}}\right) \tag{17}$$

We note that the existence of the two dimensionless parameters, Reynolds number Re and relative roughness $\varepsilon/D$, makes the expressions of different partitions relatively simple, and they are independent and uncoupled. Inspired by this, we can still use the hierarchical symbolic regression method to extract the optimal dimensionless parameter combination. On the one hand, we hope that the optimal dimensionless parameter combination can maintain higher prediction accuracy and generalization ability; on the other hand, we also hope that the optimal dimensionless parameter combination can make the highest order of the outer polynomial lower, which means that the complexity of the overall model is lower, indicating that its physical interpretation is stronger. Therefore, the optimal dimensionless parameter combination is a trade-off between prediction accuracy and complexity (polynomial order). Therefore, we can screen out the optimal dimensionless parameter combination by comparing the loss under different polynomial orders.

Referring to the work of DimensionNet [21], we select the following basis vectors from many dimensional index matrices $w_b$, namely

$$w_{b1} = [1,-1,0,1]^T$$
$$w_{b2} = [2,-2,1,1]^T$$

So the causal expression (13) can be written as

$$f_D = f(s_1, s_2) = f(\frac{V\varepsilon}{\nu}, \frac{V^2 \varepsilon D}{\nu^2}) \tag{19}$$

Next, we use the experimental data collected by Nikuradse to "rediscover" the recognized valid dimensionless numbers, taking $\mathbf{s} = (s_1, s_2)$ as the input parameter and $\log(100 f_D)$ as the output parameter to extract the optimal and most physically interpretable parameter combination. In **Figure 7**, we test the fitting error Loss under different polynomial orders, and select the optimal dimensionless parameter combination by measuring the change in the relationship between Loss and order. We can find that below the 4th order, as the order increases, the magnitude of the fitting error decreases rapidly, and it is in an underfitting state; above the 4th order, as the order increases, the magnitude of the fitting error does not change much, and it enters an overfitting state. Therefore, the dimensionless parameters obtained by the 4th order are considered to be the result of a trade-off between accuracy and complexity. From the table of parameter combinations corresponding to different orders in **Figure 7**, we can find that the dimensionless parameters obtained by the fourth order are Reynolds number $Re = s_2/s_1$ and relative roughness $\varepsilon^* = \varepsilon/D = s_1^2/s_2$. At the same time, we find that the Reynolds number Re and relative roughness $\varepsilon/D$ appear very frequently in the table, which indirectly reflects that this parameter combination is the best. This chart proves from a data-driven perspective that Reynolds number Re and relative roughness $\varepsilon/D$ are the best dimensionless parameter combination to describe the rough circular pipe flow system. Finally, we selected 80% of the data as the training set and 20% of the data as the test set. The model obtained from the training set was verified on the test set to obtain the prediction error and determination coefficient $R^2$ of Reynolds number Re and relative roughness $\varepsilon/D$. At the same time, the sensitivity analysis was also performed. The Reynolds number Re is more important because it controls the basic state of the flow and the friction factor in most areas. Relative roughness is only the only variable in specific areas.

The work we refer to is DimenisonNet [21]. Although the two methods have different practical processes, the concepts are the same and the results are consistent. The Hi-$\pi$ method changes the model complexity by changing the order of the polynomial, and makes a trade-off between the order and accuracy. DimensionNet balances the relationship between the accuracy and complexity of the model by changing the BIC threshold, thereby selecting simpler parameters. However, our method is a white box model with simple operation and controllable process. Compared with DimenisonNet, it does not require more suitable hyperparameters and optimization algorithms. This method uses a limited data set and a data-driven method to find the optimal and most physically interpretable dimensionless parameter combination for the complex system, thereby achieving higher prediction results while improving the interpretability of physical analysis.

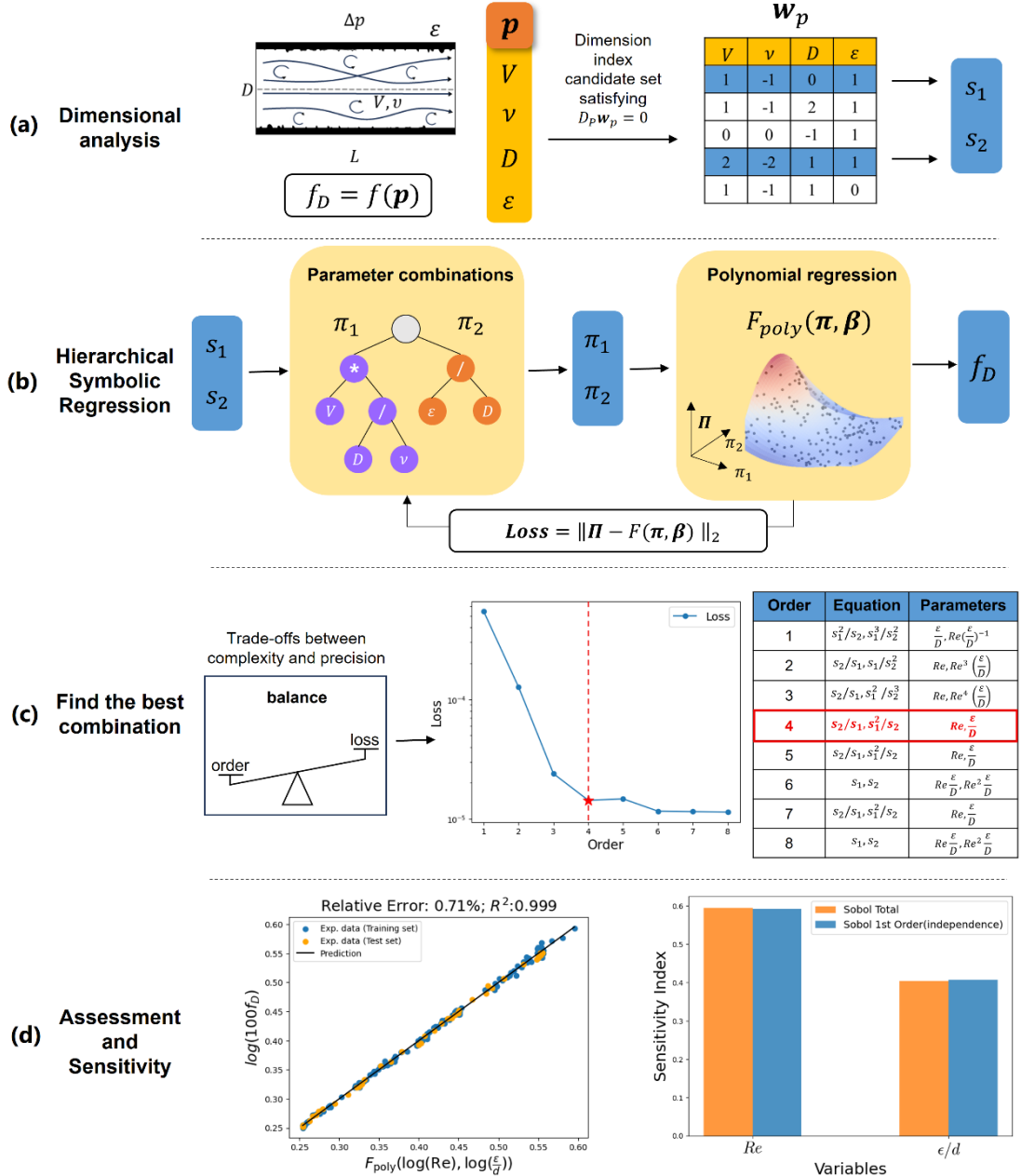

**Figure 7 Schematic diagram of optimal parameter combination extraction. (a)** Determine the parameter relationship, perform dimensional analysis, and then select any set of dimensionless parameter combinations $s = (s_1, s_2)$. You can also select other dimensional index matrices $w_b$, and the final result is the same; **(b) (c)** Then use the hierarchical symbolic regression method to extract parameter combinations, scan from 1st to 8th order, test the corresponding fitting errors and extracted parameter combinations under different polynomial orders, and finally determine that the 4th order corresponding parameter combination is the best and most physically interpretable according to the trade-off between complexity and accuracy; **(d)** Finally, compare the predictive ability and sensitivity of the dimensionless parameters.

## 3.4 Compressibility correction in subsonic flow

In aerodynamic research, compressibility correction is the key to high-speed aerodynamic design, which can improve calculation accuracy, optimize performance and reduce test costs. The pressure coefficient $C_p$ is a dimensionless parameter describing the fluid pressure distribution, and the Mach number $M$ is a dimensionless parameter measuring the degree of air compression.

The pressure coefficient is defined as

$$C_p = \frac{p-p_\infty}{\frac{1}{2}\rho_\infty V_\infty^2} \tag{20}$$

The Mach number is defined as

$$M = \frac{V}{c} \tag{21}$$

where $p_\infty$ is the freeflow static pressure, $\rho_\infty$ is the freeflow density, $V_\infty$ is the freeflow velocity, $p$ is the local static pressure, $V$ is the local velocity, and $c$ is the local sound speed.

When M<0.3, the flow is incompressible, and the pressure coefficient (i.e., the pressure coefficient of incompressible flow $C_{p,0}$) can be expressed as

$$C_{p,0} = 1 - \left(\frac{V}{V_\infty}\right)^2 \tag{22}$$

If M>0.3, the flow is in a compressible state, and the compressibility effect cannot be ignored. Therefore, it is necessary to make a compressibility correction to $C_{p,0}$ to estimate the pressure coefficient of compressible flow. $C_p$ is expressed as a function of the pressure coefficient of incompressible flow $C_{p,0}$ and Mach number $M$, which can quantitatively describe the effect of compressibility on pressure distribution, i.e.

$$C_p = f(C_{p,0}, M) \tag{23}$$

The expressions of this relationship are all dimensionless quantities, and there is no need to perform physical dimensionality reduction through dimensional analysis.

Common compressibility correction formulas include
- Prandtl-Glauert correction:

$$C_{p,PG} = \frac{C_{p,0}}{\sqrt{1-M^2}}; \tag{24}$$

- Karman-Tsien correction:

$$C_{p,KT} = \frac{C_{p,0}}{\sqrt{1-M^2}+[M^2/(1+\sqrt{1-M^2})]C_{p,0}/2}, \tag{25}$$

where the Prandtl-Glauert correction formula is applicable to two-dimensional subsonic flows with low freestream Mach numbers and close to the freestream pressure region; the Karman-Tsien correction formula is applicable to two-dimensional subsonic flows with medium and low freestream Mach numbers, taking into account nonlinear effects [36,37].

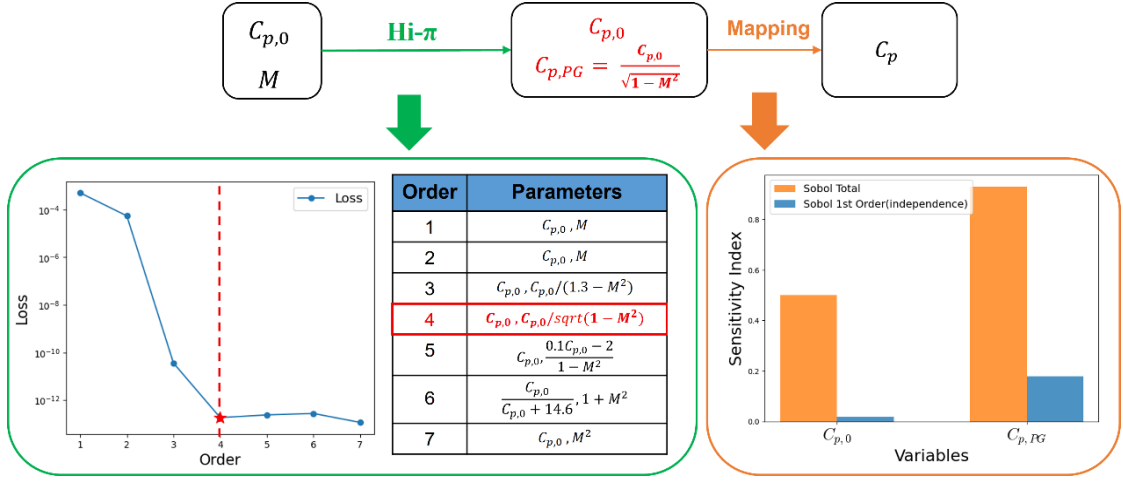

**Figure 8 Parameter combinations extracted from the compressibility correction data set.** The pressure coefficient $C_p$ of the compressible flow may be the joint effect of the incompressible case and the Prandtl-Glauert correction formula at low Mach numbers. According to the sensitivity analysis, it can be concluded that both parameter combinations are important.

Next, we use the more complex Karman-Tsien compressibility correction formula to construct a synthetic data set, generate 100 sample points in the range of $C_{p,0} \in [-1,1], M \in [0,0.7]$ through Latin hypercube sampling, add noise, and use the Hi-$\pi$ method to extract the dimensionless parameter combination in the data set. As shown in **Figure 8**, we test the fitting loss and the corresponding parameter combination under different polynomial orders. We can find that below the 4th order, as the order increases, the magnitude of the fitting error decreases rapidly, and it is in an underfitting state; above the 4th order, as the order increases, the magnitude of the fitting error does not change much, and it enters an overfitting state. Therefore, according to the order-loss curve, the result of the 4th order extraction is selected, that is, the pressure coefficient $C_{p,0}$ of the incompressible flow and the Prandtl-Glauert correction formula $C_{p,PG} = \frac{C_{p,0}}{\sqrt{1-M^2}}$. This parameter combination $C_{p,PG}$ is a complex nonlinear parameter combination and cannot be expressed in the form of a power law. Therefore, equation (23) can be rewritten as

$$C_p = f\left(C_{p,0}, \frac{C_{p,0}}{\sqrt{1-M^2}}\right) \qquad (26)$$

Through the above formula, we find that the pressure coefficient $C_p$ of compressible flow may be the combined effect of the incompressible solution and the compressibility correction solution under the Prandtl-Glauert small perturbation theory. In simple cases, the first term (incompressible case) or the second term (i.e., Prandtl-Glauert correction) may be directly taken; but in more complex scenarios (such as high Mach numbers or nonlinear effects), a more complex combination may be required. At the same time, the process of parameter combination extraction also reveals that the Prandtl-Glauert correction formula is included in the Karman-Tsien correction formula, further proving the advantages of hierarchical symbolic regression for optimal parameter combination extraction and knowledge discovery. At the same time, we conducted a sensitivity analysis on the two parameters $\left(C_{p,0}, \frac{C_{p,0}}{\sqrt{1-M^2}}\right)$, further illustrating that both parameters have an important influence and act together on the pressure coefficient, and the cross-term influence of the two accounts for a large proportion.

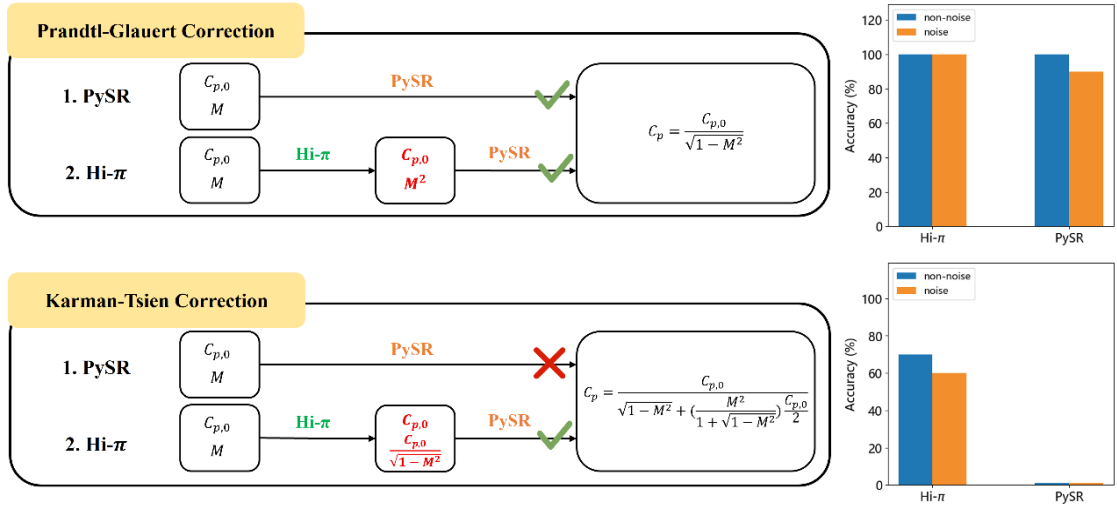

**Figure 9 Comparison of two methods in knowledge discovery of compressibility correction formula.** The upper figure is Prandtl-Glauert correction formula discovery; the lower figure is Karman-Tsien correction formula discovery.

The various examples above have proved the advantages of extracting the optimal parameter combination by the hierarchical symbolic regression method. This section applies the advantage of optimal parameter combination extraction to the direction of symbolic regression knowledge discovery, solving the previous knowledge discovery problem of complex mathematical structure formulas. To this end, we define two ideas (see **Figure 9**):

- **PySR**: directly use PySR for symbolic regression knowledge discovery;
- **Hi-$\pi$**: first use the hierarchical symbolic regression framework to extract the optimal parameter combination of the input variables, and then use PySR for symbolic regression knowledge discovery using the transformed variables as input.

In the Prandtl-Glauert example, the final results of the two methods have the same accuracy. Since the mathematical structure of the Prandtl-Glauert correction formula is relatively simple, the effect of Hi-$\pi$'s optimal parameter combination extraction is not significant. First, the optimal parameter combination of the original variable $(C_{p,0}, M)$ is extracted, and the result is $(C_{p,0}, M^2)$, and then symbolic regression knowledge discovery is performed. After adding random noise, the accuracy of Hi-$\pi$ is slightly higher than that of the PySR method.

In the Karman-Tsien example, the PySR method directly performs multiple symbolic regressions from the original variables. Due to the complex mathematical structure of the Karman-Tsien formula, the PySR method fails. By analyzing the results, we found that the symbolic regression method cannot use the low-dimensional manifold characteristics of the data set to construct and search for suitable individuals. It tends to converge to single-layer structure expressions such as integers and is insensitive to fractional expressions such as the Prandtl-Glauert correction formula. Among them, the integer form $C_p = C_{p,0} + C_{p,0}M^x f(C_{p,0}, M)$ appears the most times, resulting in a low accuracy rate of the results. However, when we use the Hi-$\pi$ method for knowledge discovery, the accuracy is still 70%. The Hi-$\pi$ method first extracts the optimal parameter combination of the original variable $(C_{p,0}, M)$, and the result is $\frac{C_{p,0}}{\sqrt{1-M^2}}$, and then performs symbolic regression knowledge discovery. The accuracy is still very high. Under the condition of adding certain random noise, the symbolic expression discovery of complex structures remains robust.

The mathematical structure of the Karman-Tsien correction formula is very complex. When facing this kind of complex structure formula, the traditional symbolic regression extraction process is like looking for a needle in a haystack. The extraction accuracy is poor and the robustness is low. We propose a new idea – first extract the optimal parameter combination, and then use symbolic regression. We found that the success rate of extracting complex structure expressions is greatly increased. The idea of using parameter transformation and then symbolic regression was also mentioned in the research work of AI-Feynman [38] and was regarded as an important step in its process (equivalent to the thinking process before symbolic regression). That is, by appropriately transforming the dependent and independent variables, the probability of successfully discovering the equation expression can be effectively improved. However, it should be pointed out that in AI-Feynman, these transformations are not automatically completed by the algorithm, but a set of predefined operations designed based on the researcher's domain knowledge and experience. The algorithm only searches for the best combination among these predefined operations. The Hi-$\pi$ algorithm we proposed uses a data-driven approach to automatically discover the optimal parameter combination. By finding the optimal substitutions and transformations, the possibility of falling into a local optimal solution during the search process may be reduced, significantly increasing the probability of mining accurate expressions from the data.

# 4. Conclusion

Dimensional analysis is a physical dimension reduction method based on dimensional invariance. On this basis, we propose Hi-π, a hierarchical dimensionless learning method driven by physics (dimensional analysis) and data (symbolic regression and polynomial regression) to extract parameter combinations that have a key impact on the output. We show that this method can accurately capture the intrinsic dimensions in high-dimensional data, achieve a balance between model accuracy and complexity, and improve the accuracy and generalization of physical modeling through mathematical and physical examples. Finally, the Hi-π method also helps to overcome the shortcomings of traditional symbolic regression in capturing hierarchical expressions and improves the ability of data-driven knowledge discovery.

The advantages of our method are summarized as follows:

- **Flexibility**: Symbolic regression allows for high flexibility in the extracted parameter combinations without the pre-assumption that the parameter combinations are in a fixed form of power multiplication.
- **Comprehensiveness**: PySR is based on a multi-branch tree structure. This design enables the model to expand from a single parameter combination to discover multiple parameter combinations, thereby more comprehensively mining potential relationships.
- **Simplicity**: Compared with neural network methods, the Hi-π method has the advantages of simple structure, interpretability.
- **Robustness**: This method shows strong robustness in high parameter dimensions, sensitivity to different parameter combinations, and noisy data.

It is also necessary to point out several limitations of this method:

- **Scarce samples**: The sample data volume and accuracy required by the multiple parameter combinations extraction method are higher than those of a single parameter combination extraction method;
- **Complex mapping relationship**: The ability of polynomial regression to represent mapping relationships may be relatively limited in high-dimensional complex problems, and model-free methods of information theory can be used to explore.

In short, hierarchical dimensionless learning (Hi-π) method provides an effective research framework for low-dimensional representation of high-dimensional complex systems. In the future, we hope that more such methods will be proposed and the table in **Figure 1** will be continuously improved. We also hope that it can be more widely used in other scientific fields.

**Author contributions**

Mingkun Xia: Methodology, Validation, Visualization, Writing-original draft, Writing-review&editing; Haitao Lin: Investigation, Methodology, Validation; Weiwei Zhang: Conceptualization, Funding acquisition, Investigation, Methodology, Project administration.

**Competing interests**

There is no conflict of interest between the authors and other persons organizations.

**Data availability**

Data will be made available on request.

**Acknowledgments**

This research was supported by the National Natural Science Foundation of China (No.92152301, No. U2441211, No. U23B6009).